# Control of Computer Peripherals using Human Eyes

Pritish Urumkar
Dept. of Electronics and Telecommunication Engineering,
SVKM's NMIME, MPSTME.
Mumbai, India.
pritish.urumkar@gmail.com

Ashwini Gade
Dept. of Electronics and Telecommunication Engineering,
SVKM's NMIME, MPSTME.
Mumbai, India.
ashwini.gade@nmims.edu

*Abstract:* **The essential activities such as communication via email, surfing the world wide web, watching ones preferred Film or television series a large majority of people impaired by neurolocomotor disorders including those paralyzed by accident do not access machines. It was inferred from a previous research review those eyeballs are really an exceptional contender for pervasive computers because eyeballs switch involuntary through accordance via computer equipment. It may be important to use this underlying information from eye motions to carry the use of machines back to those patients. We suggest an Eye-Gaze Cursor control device for this function that is fully controlled only with the use of eyeballs. Goal of this project is to develop a easy to use eye-gesture control device which will detect eye movements robustly and allow the person to use a computer webcam in accordance to the behavior mirroring to specific eye movements/gestures. It distinguishes the pupil from the face of the user and then controls its gestures. In real-time, it has to be specific enough that the user can use it with ease like most daily gadgets.**

## I. Introduction

As the name suggests is a tool which allows the user to operate the cursor on a PC or Laptop, hands-free without any wearable hardware or sensors. This HCI (Human- Computer Interaction) application will allow you to control your mouse cursor with your eye movements. The framework interfaces specifically with the person's eyeballs vision and then activates the program. Eyeball's motion, a legitimate action validation framework that uses the users' eye gestures to manipulate a virtual mouse pointer. Just one condition for the mouse device to function is that users have at the very least a single eyeball enabled with sight and machine control capabilities. It can be used by adults and young people with ALS, multiple sclerosis, strokes of the brain stem, brain paralysis, spinal cord trauma, emotional wounds, and so on. The Eye-Gaze Cursor device can be used as tool for residences, businesses, long-distance mental health offices, schools, and healing centers. An individual can operate a computer mouse, access the internet, and even send and receive email by looking at the program on a device that is projected on a screen.

A substantial portion of the population, affected by neurolocomotor disorders or paralyzed by illness or injuries, is unable to use computers for minor activities such as internet communication, browsing and entertainment usage. Its shared light on an interesting insight that eyeballs are a fantastic prospect for casual computer usage through thorough research analysis, as they switch anyway during contact with computing machinery. It may be important to use this underlying information from eye motions to introduce the use of machines back to those patients. We suggest a mouse motion control device for this function, which is controlled entirely by the persons eyeballs only. Goal of this project is to encourage the user to do activities that are mapped to eye movements/gestures so that the person can easily utilize it like most daily peripherals.

Many patients are also victims of illnesses that physically damage them, such as Paraplegia, which means that a human is ineffectual to make use of his or her abilities or her the whole torso from the chin down. The only organ which can produce multiple acts is their eyes. A total of 518 million people in the 7 billion population in 2011 Census, people said they had a disability. Actually, about 10 percent (about 650 million) of people living with disabilities in the world's population reside on Feb 7, 2018.

A large number of individuals affected by or paralyzed by amyotrophic lateral sclerosis (ALS) [3] are unable to utilize machines with ordinary daily operations tasks. They need assistance from another human to feed them, even when it comes to eating. For their day-to-day affairs, these people are struggling. Disabled persons actually normally by putting long sticks in their mouths, they type on a computer keyboard. The alternative we put forth will enable disabled persons in their lives to be autonomous. It will provide them with an incentive in their lives to entertain, socialize and work. For interaction with digital instruments, existing electronic input devices, such as a mouse and keyboard, as well as other types of input devices, will be used. This electronic input machines cannot be used on their own by people with disabilities. HCI's innovative and specialized methods are developed easily. In this research sector, several experts are actively employed. Persons eyeballs have abundant of data that can be accessed that can be used in different implementation [2] (i.e., Computer interaction).

## II. Literature survey

The most significant task was to consider the area of study in which a mouse is engaged in eye recognition and cursor movement. The emphasis of the literature was on how to build a method that would accommodate the needs of people with physical disability and the framework ought to be very clear to comprehend.

A community at MIT has developed a methodology coined "The Sixth Sense," which seeks to facilitate human-computer interaction using hand and eye movements. [1] The whole structure can be mounted on the head of the user, meaning it is possible that a projection will be on flat construct (e.g., screen or panel) and can be utilized all over the globe. Concern is that it cannot offer the affected with better support and connectivity, nor does it build a framework that can communicate with other connectivity options.

While a detailed description is provided by Drewes, Heiko, it was discovered that most architectures required more improvement while it used repetitive and time-consuming calibration methods. [2] Schmidt, Jochen, identified a no-nonsense, agile method by using the motor architecture framework for human-computer interaction. William Rhoades Patera, Kassner, and Moritz Philipp exploited this by utilizing similar SFM (motion construct) algorithm, improving it and developing on its applications as an effective pupil tracking architecture. [3] In order to accomplish this purpose, a system called PUPIL was created to objectively examine the interaction with respect to a space and the human subject in order to imagine thisspatial perception and to facilitate its utilization for the tracking of eyeballs movements.

During 2014, a method was developed using an eye monitoring system focused on the collection of pictograms. [5] The problem with this device is that it would not operate if any fluid is present in the eyes. In 2016, a wireless eye-gaze detection device based on Vision was instigated.[8] This system uses a high infrared camera to operate. This sense the eyeballs of a human from the infrared cameras. The trouble with this scheme was that it was inefficient and expensive. A better system was launched in 2017; this system has been developed for paralytic patients.[9] To identify the pupil, it utilized many of techniques and algorithms. In 2018, a Hough transform-based eye tracking algorithm was developed. [10] This device senses a person's face and eyes. In this framework, the challenge is real-time monitoring and time-speed. We are proposing a method using "Facial Landmark System" which is predominantly use for overlaying filters, marks, animation etc. on a human face. A combination of these system along with using modules like Hough Transform, Harr Cascade, mean shift algorithm, Kalman Filtering etc. we can have a robust, reliable, cheap and efficient system for developing Eye Gaze Cursor.

### III. Algorithm

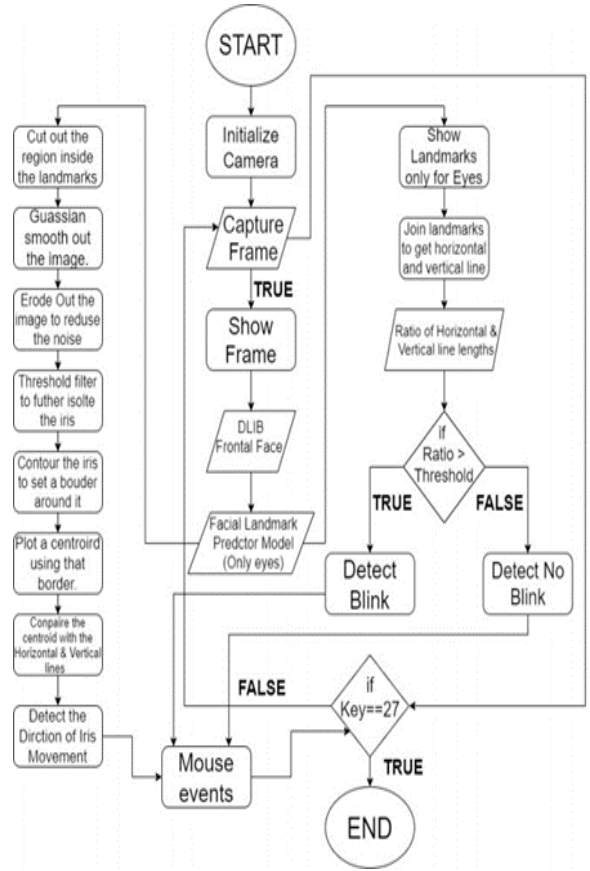

**Figure 1: Algorithm of the complete workflow of the system**

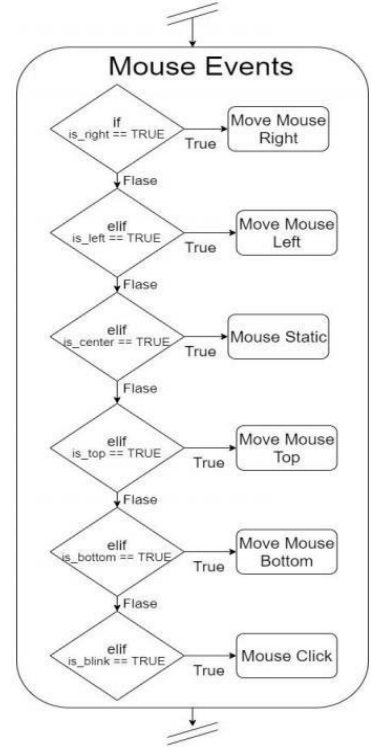

**Figure 2: Algorithm of the integration of mouse events**

## IV. Method

### A. Getting Video Feed from Webcam

A video from a webcam is obtained, using Python and OpenCV. Then, the video is displayed on the OpenCV window. A live stream is captured with the employment of webcam. To capture a video from the webcam, OpenCV offers a very basic interface.

### B. Detecting Face in the Video Feed

A video from a webcam, using Python and OpenCV is obtained and a human face in that video frame is detected. The video is played along with the rectangular grid around the detected face on an OpenCV window. A live stream is captured with a camera that has a grid that detects face in each of the frames from this live stream video feed. To detect a human frontal face in a frame, an object of class needs to be created which is available in the **dlib** library. The class is trained for detecting a human face, so we can simply use it for this purpose.

### C. Detecting Eye in the Video Feed

A live stream is captured with a camera that has a grid that detects a face as well as the eyes in each of the frames from this live stream video feed. The facial landmarks points are detected, to detect human facial features in a frame, as we need to create an object class shape predictor which is available in the **dlib** library. As input, the constructor of this class receives a **pre-trained module of facial landmarks** that is trained using a machine learning module available in the **dlib** library. The class is trained for detecting a human face, so we can simply use it for this purpose. This facial landmark module assigns a landmark assigning a number to every facial feature on the face. We just use the landmarks around the eyes indoor to isolate the detection of the eyes in the live video feed.

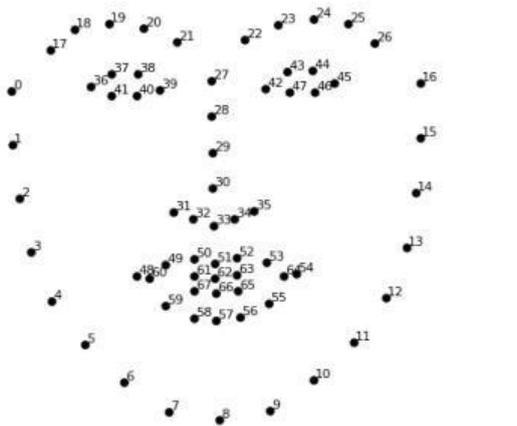

**Figure 3: Pretrained facial landmark module in the DLIB library**

### D. Eye blink Detection in the Video Feed

A live stream with a camera that has a grid is captured that detects the eyes and its blinks in each of the frames from this live stream video feed. To detect the blinks, the landmarks traced around the eye are used that we achieved in order to locate the eyes in the live video feed. Now, using those land marks we will create a horizontal and vertical lines, in order to detect the blinks of the eyes. To predict the blinks, the ratio of the horizontal and vertical line will be calculated, and then compared with a threshold vale to predict the blinks. We, then this ratio against a threshold value will be used to detect the blinks.

### E. Iris Detection in the Video Feed

Now, as just the eye in the frame is isolated, it will be processed through various filters in order to detect the pupil accurately using modules and filters in the OpenCV library. The frames coming from the live video feed which just have the cut- out part of the isolated eye region will be treated just as images which will be considered as the eye frame. Firstly, the image through a gaussian filter known as the Bilateral Filter will be passed which will blur the image to get rid of all the depths, textures as well as any features in the images which we will save the results in. Secondly, the image through a filter known as the Erode Filter will be passed which will erode the image to get rid of unwanted area around the eye as well as reduce any white noise present in the images which we will save the results in new frame which will act as a holding directory. Thirdly, the image through a filter known as the Threshold Filter will be passed which will further remove any noise from the images which we will save the results in eye frame which will act as a holding directory. After processing the fame through various image processing filters from the OpenCV library, finally the pupil from the eye will be isolated. But as the pupils shape is very random along with edges and curves not fitting into any Definity proper geometric shape. To solve this, we a module from the OpenCV library known as the Find Contours will be used, which will draw a contour around the pupil in the frame giving it a definite geometric shape. Finally, after locating as well as isolating the pupil in a definite geometric shape, we can plot a marker in the middle on the pupil in order to get the exact position of the pupil with the help of finding the centroid which in turn will be the marker.

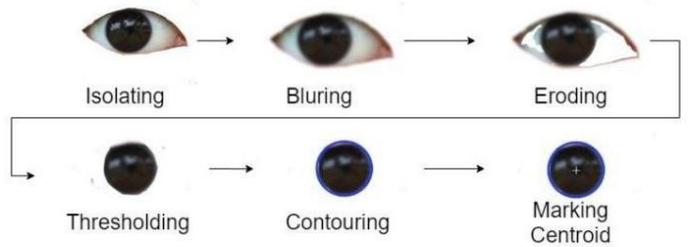

**Figure 4: Transition of using filters for the detection of Iris**

Integration of Eye with Mouse events

To trigger the mouse events, we will use the detections and tracking of the eyes as inputs. Each of the movement and gestures committed by the eyes of the user will be specifically used as input for specific mouse events. Mouse events will be resourced from the **PyAutoGUi** module which will give the mouse events like movement of the cursor as well as the clicks.

## V. Results

To follow the movement of pupil we have marked a centroid on the pupil which can be used to track the exact position as well as the direction of the moving pupil. For this we will again make use of the Landmarks which are been plotted around the eye using the **Dlib** module. By joining the horizontal corner landmark, we get the horizontal line with a definite length and joining the top bottom landmarks we get the vertical line with a definite length, the same way did to find out the ration to determine the blinks. As there are two eyes, the two of the horizontal line are used to calculate the horizontal ration and two vertical line to calculate the vertical ratio. Then the centroid with the horizontal ratio and vertical ratio are compared in order to get a position as well the direction of the movement if the pupil by using the ratios as the thresholds. We create functions for each of the direction for the ease of use while integration with the mouse events.

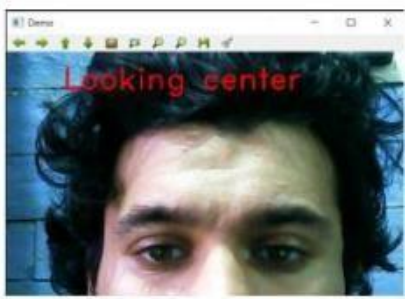
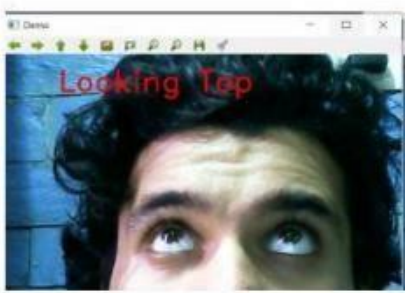

**Figure 5 (A)**

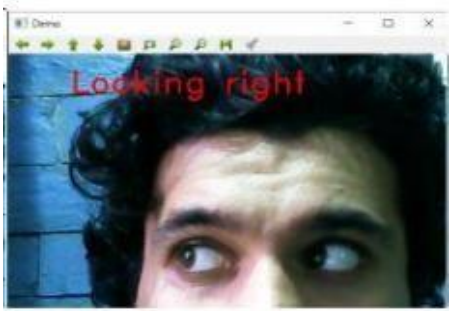
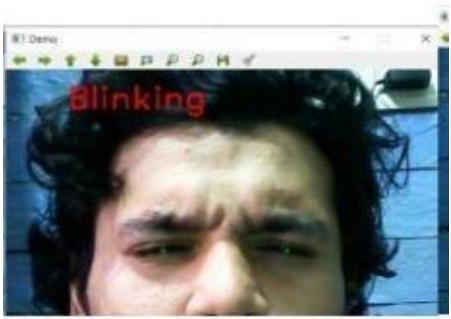

**Figure 5 (B)**

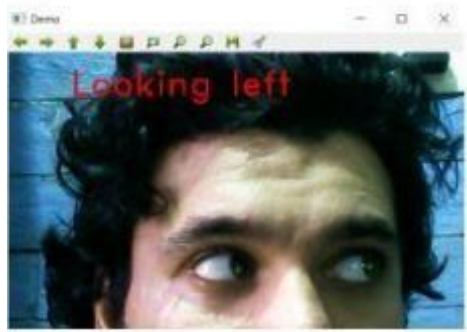
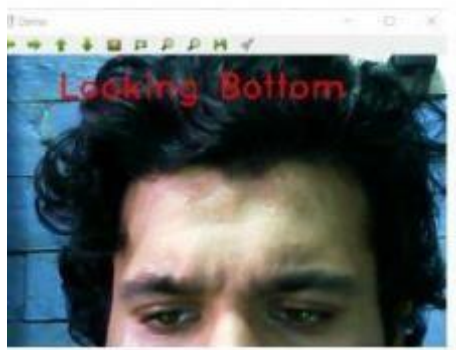

**Figure 5 (C)**
**Figure 5: Result of the system**

VI. Conclusion

A framework that empowers a crippled individual to communicate with the PC was effectively evolved and tried. The strategy can be additionally improved to be utilized in numerous different applications. The framework can be adjusted to assist the incapacitated with controlling home machines, for example, TV sets, lights, entryways and so forth. The framework can likewise be adjusted to be utilized by people experiencing total loss of motion, to work and control a wheelchair. The system can likewise be utilized to recognize laziness of drivers so as to forestall vehicle mishaps. The eye development recognition what's more, following have additionally potential use in gaming and computer-generated reality.

**REFERENCE**

[1] Heiko Drewes and Albrecht Schmidt. "Interacting with Computers Using Gaze Gestures" Human-Computer Interaction - INTERACT 2007, 11th IFIP TC 13 International Conference, Rio de Janeiro, Brazil, September 10-14, 2007, DOI: 10.1007/978-3-540-74800-7_43.

[2] PUPIL: constructing the space of visual attention. Diss. Massachusetts Institute of Technology, 2012.) Kassner, Phillip, and Patera (2012)

[3] A. Lopez , P.J. Arevalo , Francisco Javier Ferrero Martin, Marta Valledor, "EOG-based system for mouse control", 2014 IEEE Conference on Sensors, DOI: 10.1109/ICSENS.2014.6985240.

[4] Chairat Kraichan and Suree Pumrin "Face and eye tracking for controlling computer functions" 2014 11th International Conference on Electrical Engineering/Electronics, Computer, Telecommunications and Information Technology (ECTI-CON), DOI: 10.1109/ECTICon.2014.6839834.

[5] Radu Gabriel Bozomitu;Alexandru Păsărică;Vlad Cehan;Cristian Rotariu and Constantin Barabaşa "Pupil center coordinates detection using the circular Hough transform technique" 2015 38th International Spring Seminar on Electronics Technology (ISSE), DOI: 10.1109/ISSE.2015.7248041.

[6] A. Păsărică;R. G. Bozomitu;V. Cehan;R. G. Lupu and C. Rotariu , "Pupil detection algorithms for eye tracking applications", 2015 IEEE 21st International Symposium for Design and Technology in Electronic Packaging (SIITME), DOI: 10.1109/SIITME.2015.7342317.

[7] A. López, D. Fernandez, F. F. Martín, M. Valledor, and O. Postolache , "EOG signal processing module for medical assistive systems upload, 2016 IEEE International Symposium on Medical Measurements and Applications (MeMeA), DOI:10.1109/MeMeA.2016.7533704.

[8] Aditya Dave and C. Aishwarya Lekshmi "Eye-ball tracking system for motor-free control of mouse pointer", 2017 International Conference on Wireless Communications, Signal Processing and Networking (WiSPNET), DOI: 10.1109/WiSPNET.2017.8299921.

[9] Metin Yildiz and Muhammet Yorulmaz, "Eye gaze location detection based on iris tracking with web camera", 2018 26th Signal Processing and Communications Applications Conference (SIU), DOI: 10.1109/SIU.2018.8404314.

[10] Michelle Alva;Neil Castellino;Rashmi Deshpande;Kavita and Sonawane;Monalisa Lopes "An image-based eye controlled assistive system for paralytic patients", 2017 2nd International Conference on Communication Systems, Computing and IT Applications (CSCITA), DOI: 10.1109/CSCITA.2017.8066549.

[11] Aleksei Bukhalov and Viktoriia Chafonova "An eye tracking algorithm based on Hough transform", 2018 International Symposium on Consumer Technologies (ISCT), DOI: 10.1109/ISCE.2018.84